# Coordinated Planning of Offshore Charging Stations and Electrified Ships: A Case Study on Shanghai-Busan Maritime Route

Hao Li, Hanqi Tao, Wentao Huang, Hongcai Zhang, Ran Li

*Abstract*—Despite the success of electric vehicles on land, electrification of maritime ships is challenged by the dilemma of range anxiety and cargo-carrying capacity. The longer range requires larger batteries, which inevitably eat up the precious cargo space and weight. This paper breaks new ground by proposing a coordinated planning model for offshore charging stations (OCSs) and electric ships (ESs), marking a first in this field. Strategically situated OCS can partition a long maritime route into several shorter segments, which in turn lead to smaller batteries and thus larger cargo capacities. The research analyzed the impact of maritime geographical conditions on the placement and sizing process and provided insights into the trade-offs between battery size, cargo-carrying capacity, and the cruising range of different types of electrified ships. Using real Automatic Identification System (AIS) data, we estimated the economic feasibility of the Shanghai-Busan high-traffic maritime route and conducted a sensitivity analysis on factors affecting its economic viability. The results show that installing OCS can significantly reduce the propulsion cost compared with ESs without OCS and traditional internal combustion engine (ICE) ships.

*Index Terms*—Offshore Charging Station, Electric ships, Coordinated Planning.

## NOMENCLATURE

*Indices and Sets*

| | |
|---|---|
| $i$ | Index of OCS candidate points |
| $t$ | Index of time periods |
| $v$ | Index of ships |
| $s$ | Number of operation scenarios |
| $D_s$ | Set of geographical factors at scenario $s$ |
| $G$ | Set of shipping traffic volume |
| $S$ | Set of scenarios |
| $\mathcal{J}$ | Time scale |
| $I$ | Set of OCS |

*Parameters*

| | |
|---|---|
| $\mu_{reg}^{elec}$ | Price per Unit onshore electricity price in different ports |
| $\mu_{batt}$ | Price per Unit capacity battery |
| $\mu_{carg}$ | Revenue per unit volume of cargo |
| $V_b, V_c$ | Battery and cargo volume per unit capacity |
| $V_0$ | Total available volume of the cabin |
| $\mu_{res}, \mu_{BESS}, \mu_{char}$ | Cost of renewable energy source (RES), battery energy storage systems (BESS) and charging infrastructure per unit installed capacity |
| $k_w, k_{PV}$ | Power generation coefficients of wind and photovoltaic |
| $k$ | Coefficient of draught |
| $C_0^{plat,fixed}, C_0^{plat,float}$ | Cost Benchmark of fixed platform and floating platform |
| $\rho_W$ | Density of ocean water |
| $\rho_b, \rho_c$ | Weight coefficient of battery and cargo |
| $L_{rt}$ | Route lengh |
| $S_v$ | Projected area of the deck |
| $\mu_d^{fixed}$ | Cost coefficient of water depth for fixed platforms |
| $\mu_d^{float}, \mu_v^{float}, \mu_h^{float}$ | Cost coefficient of water depth, wind speed and wave height for floating platforms |
| $v_i$ | Wind speed at candidate point $i$ |
| $d_i$ | Ocean water depth at candidate point $i$ |
| $h_i$ | Wave height at candidate point $i$ |
| $R_i$ | Solar radiation intensity at candidate points $i$ |
| $w_s^{scn}$ | Weights for each scenario |
| $W$ | Number of weeks throughout the entire life cycle |

*Variables*

Hao Li and Hanqi Tao are with the School of Smart Energy, Shanghai Jiao Tong University, Shanghai 200240,China (e-mail: lihao1992@sjtu.edu.cn, johnappleseed@sjtu.edu.cn)
Wentao Huang(Corresponding author) is with the Key Laboratory of Control of Power Transmission and Conversion, Ministry of Education, Shanghai Jiao Tong University, Shanghai, China (e-mail: hwt8989@sjtu.edu.cn)
Hongcai Zhang is with the State Key Laboratory of Internet of Things for Smart City, Zhuhai UM Science and Technology Research Institute, Macao, China (e-mail: hczhang@um.edu.mo)
Ran Li(Corresponding author) is with the Non-Carbon Energy Conversion and Utilization Institute, Shanghai Jiao Tong University, Shanghai 200240, China (e-mail: rl272@sjtu.edu.cn)

| | |
|---|---|
| $x_i$ | Binary variable for whether install OCS at candidate point $i$ |
| $C_{apc}$ | Average propulsion cost (APC) |
| $C_{ocs}$ | Cost of OCS |
| $C_{elec}^{es}, C_{batt}^{es}, C_{carg}^{es}$ | Cost of onshore electricity, ship battery and revenue from cargo |
| $Q_i^{res}$ | RES installed capacity of point $i$ OCS |
| $Q_i^{BESS}$ | BESS installed capacity of point $i$ OCS |
| $Q_i^{char}$ | Charging infrastructure installed capacity of point $i$ |
| $Q_{es}, Q_v^{es}$ | Battery size, battery size of ship $v$ |
| $\delta_{es}, \delta_v^{es}$ | Cargo-carrying capacity, cargo-carrying capacity of ship $v$ |
| $L_v^{es}$ | Cruising range of ship $v$ |
| $P_{i,t}^{load}$ | Charging power of load at point $i$ at time $t$ |
| $P_{i,t}^{res}, P_{i,t}^{BESS}$ | Output power of RES and BESS at point $i$ at time $t$ |
| $P_i^{UB}$ | Maximum charging power of OCS $i$ |
| $Q_{i,t}^{BESS}$ | Capacity of BESS at point $i$ at time $t$ |
| $V_{ICE}, V_e$ | Volume of ICE propulsion system and electric propulsion system |
| $T_0$ | Draught |
| $W_{ICE}, W_e$ | Weight of ICE propulsion system and electric propulsion system |
| $n_w, n_{PV}$ | Number of wind turbines and photovoltaic units |
| $\alpha_{depr,i}^{ocs}, \alpha_{depr,v}^{es}$ | Depreciation Coefficients for OCS and ES |

## I. INTRODUCTION

Maritime transport serves as a crucial pillar of the global economy, carrying over 90% of international trade goods[1][2]. It plays a unique and significant role in global trade with its characteristic of large cargo-carrying capacity and extensive voyage distance range from 1400 to 1500 kilometers. Traditional internal combustion engine (ICE) ships run on Heavy Fuel Oil (HFO) or Very Low Sulfur Fuel Oil (VLSFO) to meet this cruising range requirement with the cost of carbon emission. Following the IMO GHG Emissions Strategy [3] reducing the environmental impact of maritime shipping by transitioning to electric-powered ships has become an emerging trend.

However, a critical hurdle in the development of ESs is the dilemma between cruising range and cargo-carrying capacity. Fully ESs, despite their emission reduction, face limitations on cargo-carrying capacity due to the large battery size required to fulfil the voyage distance, considering today's lithium-ion battery energy density of 300 Wh/kg. If the cargo loss is offset with smaller batteries, the single-charge range will be only between 80-200km, insufficient for oceanic navigation. Even with projected improvements in battery energy density to 1200Wh/L by 2050, they are unlikely to meet the range and cargo-carrying capacity requirements simultaneously[4].

Mirroring the land solution of electric vehicles (EVs), a novel idea is to develop OCSs for ESs. As shown in Fig.1, the OCS utilizes offshore renewable energy to charge ESs through a buoy or a platform on the sea. Since 2020, rapid developments have been observed in OCS research projects. For instance, Maersk and Örsted[5][6] began planning to test floating OCSs in 2020. In 2023, MJR Power and Automation in the UK installed the world's first OCS between Lynn and Inner Dowsing wind farms[7], and conducted testing of interconnections, mooring, automation, monitoring, and safety systems, with preparations for further installation at offshore wind farms in the North Sea.

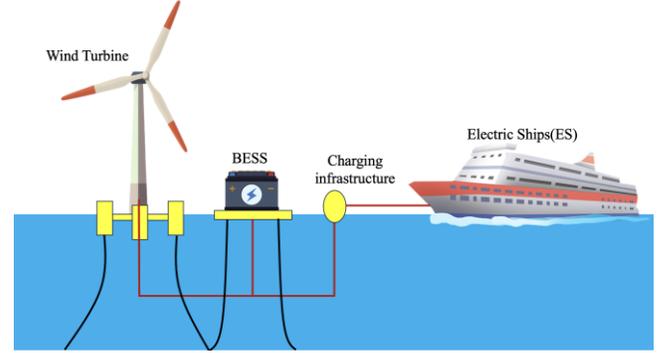

**Fig.1** Conceptual diagram of OCS.

There have been scattered studies on offshore charging platforms. These studies are mainly focused on the technical and economic feasibility study of the OCS. The operational modes of OCSs and solutions for the connectivity of ESs with OCS facilities is discussed in[8]. The economic viability of ESs relative to fuel-powered ships are discussed in [9][10] using wind energy, solar power, and floating nuclear power as sources for OCSs. Emission reduction potential is analyzed in [11] with a case study on ships operating on the Mumbai to Dubai route. [12]undertakes similar studies for the North Sea to assess the feasibility of OCSs and proposes an energy management strategy to reduce reliance on backup power sources.

Yet, a significant research gap is that existing studies focus on singular platform designs instead of strategic placement of charging stations and the corresponding design of ESs. Their delicate relationship is that strategically situated charging stations can partition a long maritime route into several shorter segments. Consequently, the more charging stations are placed, the shorter travel segments will be cut for ships, which in turn lead to smaller batteries and thus larger cargo capacities. However, this approach also scales the associated costs of constructing these offshore stations, thereby striking a balance between station placement, ship battery dimensions, and cargo-carrying capacity. Besides, unlike onshore scenarios[13]-[15], the planning of OCSs is influenced by marine geographical conditions, such as water depth, wave height, shipping traffic volume, wind, and solar resources, significantly complicating site selection.

This paper, for the first time, extends the problem from land to ocean and proposes a coordinated planning model for OCSs and ESs. The first contribution is to provide a cost-analysis framework for OCSs considering the impact of geographical factors and shipping traffic volume based on real data. The second contribution is to analytically investigates the trade-off between battery size, cargo-carrying capacity, and cruising range for electrified ships, considering their hydrodynamics and propulsion demand patterns. Using real AIS and Geographic Information System (GIS) data, this paper evaluates the optimal placement of offshore charging platforms and ES design for the Shanghai-Busan maritime route. A sensitivity analysis is undertaken to investigate the impact of uncertain parameters such as ship types, offshore renewable cost, and shipping traffic volume.

The rest of the paper is organized as follows. Section II introduces the coordinated planning model and highlights two technical challenges of geographical impacts and capacity-cruising range trade-off. Section III analyzes the impact of geographical factors and shipping traffic volume on planning costs. Section IV introduces the trade-off approach between cruising range, battery size and cargo-carrying capacity. Section V conducts a case study on the Shanghai to Busan route and performs sensitivity analyses on uncertain parameters and Section VI draws the conclusion.

## II. COORDINATED PLANNING MODEL MODELING

The overarching problem is formulated with economic efficiency as the objective function, represented by the Average Propulsion Cost (APC) $C_{apc}$ per kilometer for an ES within one voyage. As shown in (2), this cost is primarily composed of four components: i) the offshore charging cost, which is mainly the construction cost of offshore renewables and charging platforms $C_{ocs}$; ii) the onshore electricity cost of the ESs $C_{elec}^{es}$ which includes the cost of cold ironing; iii) the battery cost of the ESs $C_{batt}^{es}$; and iv) the revenue $C_{carg}^{es}$ generated from the extra cargo shipping compared with HFO or VLSFO fueled ships. The breakdown of each cost is shown in (3)-(6). To coordinate the planning of ESs and OCSs, the decision variables include: i) the battery size $Q_v^{es}$ and extra cargo capacity $\delta_v^{es}$ for ESs; ii) the location of each OCS $x_i$, BESS capacity $Q_i^{BESS}$ and RES capacity $Q_i^{res}$ of each OCS.

The constraints of the optimization problem can be categorized into 5 types: RES output constraints, OCS power balance constraints, BESS power balance constraints, ES volume constraints, and cruising range constraints. The first is the RES output constraints: Constraint (7) represents that the output power $P_{i,t}^{RES}$ of RES has an upper limit $Q_i^{RES}$. Constraint (8) indicates that this upper limit $Q_i^{RES}$ depends on the local wind and solar resources represented by $G(v_i, R_i)$. In the OCS power balance constraints, constraint (9) indicates that the load power at OCS $i$ at time $t$, denoted as $P_{i,t}^{load}$, should be less than the sum of the output from RES and BESS. It should also be less than the designated maximum charging power for the OCS, denoted as $P_i^{UB}$. Constraint (10) states that the load power $P_{i,t}^{load}$ at this point is equal to the sum of the charging power of all ESs charging at this point. BESS power balance constraints indicate that there is an upper limit on the installation capacity of BESS, as described in constraint (11), and constraint (12) governs the capacity variations of BESS. ES volume constraints include constraint (13), which imposes spatial limitations on ESs. This constraint indicates that the battery volume and cargo volume must be less than or equal to the maximum available space within the ship's cabin, denoted as $V_0$. Cruising range constraint include constraint (14), which ensures that the ESs' cruising range with their batteries should be greater than the distance between any two adjacent charging stations, guaranteeing that ESs do not run out of power midway.

$$\min \ C_{apc} \tag{1}$$

$$C_{apc} = \frac{1}{L_{rt} \sum_{s \in S}(w_s^{scn} \cdot \sum_{v \in D_s} v)}(C_{ocs} + C_{elec}^{es} + C_{batt}^{es} - C_{carg}^{es}) \tag{2}$$

*where:*

$$C_i^{ocs} = \sum_{i \in I} \alpha_{depr,i}^{ocs}(C_i^{plat} + \mu_{res}Q_i^{res} + \mu_{BESS}Q_i^{BESS} + \mu_{char}Q_i^{char}) \tag{3}$$

$$C_{elec}^{es} = \sum_{s \in S}(w_s^{scn} \sum_{v \in D_s} \mu_{reg}^{elec} Q_v^{es}) \tag{4}$$

$$C_{batt}^{es} = \sum_{s \in S}(w_s^{scn} \sum_{v \in D_s} \alpha_{depr,v}^{es} \mu_{elec} Q_v^{es}) \tag{5}$$

$$C_{carg}^{es} = \sum_{s \in S}(w_s^{scn} \sum_{v \in D_s} \mu_{carg} \delta_v^{es}) \tag{6}$$

*subject to:*

$$0 \le P_{i,t}^{res} \le Q_i^{res}, \forall i \in I, t \in \mathcal{J}, s \in S \tag{7}$$

$$0 \le Q_i^{res} \le G(v_i, r_i), \forall i \in I, s \in S \tag{8}$$

$$P_{i,t}^{load} \le P_{i,t}^{res} + P_{i,t}^{BESS} \le P_i^{UB}, \forall i \in I, t \in \mathcal{J} \tag{9}$$

$$P_{i,t}^{load} = \sum_{v \in D_s} P_{i,v,t}^{Char}, \forall i \in I, v \in D_s, t \in \mathcal{J} \tag{10}$$

$$0 \le Q_{i,t}^{BESS} \le Q_i^{BESS}, \forall i \in I, v \in D_s, t \in \mathcal{J} \tag{11}$$

$$Q_{i,t}^{BESS} - P_{i,t}^{load} + P_{i,t}^{res} = Q_{i,t+1}^{BESS}, \forall i \in I, t \in \mathcal{J} \tag{12}$$

$$Q_v^{es} + V_c \delta_v^{es} \le V_0, \forall v \in D_s \tag{13}$$

$$L_v^{es} \ge \left| L_{x_i} - L_{x_{i+1}} \right|, \forall x_i, x_{i+1} = 1, i \in I, v \in D_s \tag{14}$$

The above optimization problem is facing two challenges. The first challenge is that, compared to onshore EV planning models, the planning model for OCSs will be constrained by both geographical factors $G$ and shipping traffic volume $D$. As shown in (3), the OCS cost $C_{ocs}$ can be decomposed into four parts, including the offshore platform cost $C_i^{plat}$, RES cost $\mu_{res}Q_i^{res}$, BESS cost $\mu_{BESS}Q_i^{BESS}$ and charging infrastructure cost $\mu_{char}Q_i^{char}$. Under different geographical factors like water depth and wave height, the OCS cost may vary significantly. The majority of RES cost is the installation cost of offshore renewables which is dependent on the local wind and solar resource. To balance the charging demand as show in (11-12), the cost of BESS is a function of shipping traffic volume and the compatibility between charging and renewable patterns. Additionally, higher shipping traffic volume will spread the platform cost over each charging activities, and thus lowering $C_{\text{ocs}}$.

The second challenge is the trade-off between the cruising range $L_v$, battery size $Q_v^{\text{es}}$, and cargo-carrying capacity $\delta_v^{\text{es}}$ of ESs. Battery size is determined by the cruising range, with larger battery ensuring longer voyage. Cargo revenue is determined by the cargo-carrying capacity, where a higher cargo-carrying capacity results in greater revenue. However, as shown in (13), the cargo-carrying capacity is offset by the volume of the battery as each ship has limited space. Therefore, greater cruising range results in increased battery costs and reduced cargo revenue and vice versa. Therefore, it is essential to understand the trade-off among these three factors.

### III. GEOGRAPHICAL AND TRAFFIC IMPACT ANALYSIS

Geographical factors such as water depth or renewable resources affect the platform cost of OCS. For water depth shallower than 60 meters, monopile platforms are generally used with large cylindrical steel pipes. Installing a monopile involves drilling the steel pipe into the seabed to a certain depth to provide stability. Therefore, deeper water requires the platform's foundation to be drilled into deeper underwater, leading to higher material costs. This relationship is approximately linear. Therefore, the cost of monopile platform can be calculated using the cost benchmark $C_0^{\text{plat,fixed}}$ and the depth benchmark $d_0$ through linear equation (15):

$$C_i^{\text{plat,fixed}}(G) = C_0^{\text{plat,fixed}}(d) = C_0^{\text{plat,fixed}} + \mu_d^{\text{fixed}}(d_i - d_0) \quad (15)$$

For water depth exceeding 60 meters, floating platforms are employed. These platforms can float on the water surface and are anchored in specific positions using mooring systems. For example, semi-submersible platform is supported by three buoys to keep the platform afloat and stabilized by connecting it to the seabed via cables. Tension Leg Platform (TLP) characterized by tension legs at its four corners. These tension legs are secured to the seabed via mooring cables to maintain platform stability. TLPs are typically used in deeper water areas. Equation (16) depicts the relationship between the floating-platform cost and geographical factor:

$$C_i^{\text{plat,float}}(G) = C_0^{\text{plat,float}} + \mu_d^{\text{float}}(d_i - d_0) \\ + \mu_v^{\text{float}}(v_i - v_0) + \mu_h^{\text{float}}(h_i - h_0) \quad (16)$$

Floating platforms are affected by water depth, wind speed, and wave height. In offshore areas with greater water depths, longer mooring and seabed connections are needed. Additionally, floating platforms lack a fixed base, making them less stable. Higher wind speeds can cause the platform to sway, tilt, and rotate, while higher wave heights may result in vertical movement. This requires structural reinforcements, which increases installation costs.

Geographical factors also affect the cost of offshore RES. The higher the RES endowment, the less the installed capacity is required to provide the same amount of charging capability. RES endowment refers to the local wind speed and solar radiation intensity[16]. Wind turbine power output is linearly proportional to the cubic of the wind speed $v_i$ at that point, while photovoltaic power output is linearly proportional to the solar radiation $R_i$. Assuming the power output coefficients for wind turbines and photovoltaics are represented by $k_w$ and $k_{PV}$, and the number of installed wind turbine units and photovoltaic units is denoted as $n_w$ and $n_{PV}$, then the corresponding RES output constraints are as follows:

$$Q_i^{\text{res}} \leq G(v_i, r_i) = n_w k_w v_i^3 + n_{PV} k_{PV} R_i \quad (17)$$

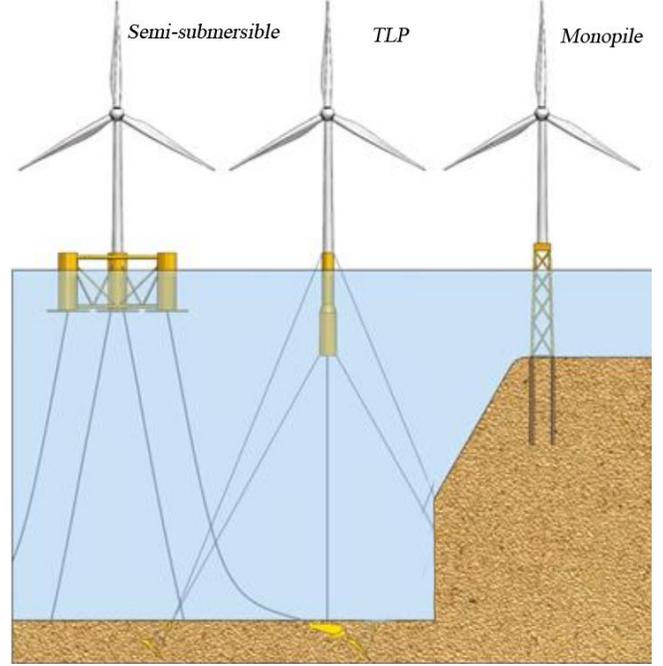

**Fig.2** Illustration of the concepts, from left to right: Semi-submersible, TLP and Monopile.

### IV. TRADE-OFF ANALYSIS OF CRUISING RANGE, BATTERY SIZE AND CARGO-CARRYING CAPACITY OF ELECTRIC SHIPS

This section will analyze the trade-off relationship between the cruising range, cargo-carrying capacity and battery size of ESs. We will first introduce the physical model of this trade-off and then, considering economic factors, establish a method to strike a balance between cruising range, cargo-carrying capacity and battery size.

*A. Physical Model of the Trade-off*

There is a non-linear relationship between battery size and cruising range. This non-linear relationship can be likened to electric vehicles (EVs). The weight of the battery affects the energy consumption of EVs, and ESs have a similar impact, but in a different manner. An increase in the weight of the ES's battery alters its own weight, subsequently increasing its draught, adding to the operational resistance of the ES, thus raising energy consumption. This relationship can be described as follows:

$$Q_{\text{es}} = kL((\Delta T + T_0)/T_0)^{\frac{2}{3}} \quad (18)$$

Equation (18) represents a nonlinear relationship between battery size and cruising range, where $k$ represents a constant based on Admiralty Law[17][18].

In order to calculate the change of draught $\Delta T$, the weight difference between the ES and ICE ship is analyzed. First is the difference between electric propulsion systems (including battery and motor, denote as $\rho_b Q_{es}/V_b$ and $W_e$) and ICE propulsion systems (including engine and oil, denotes as $W_{ICE}$). Second is the weight change in cargo, which can be represented as $\rho_c \Delta Q_c(L)$. $Q_c(L)$ is the number of containers in the unit of Twenty-feet Equivalent Unit(TEU). The weight difference can be expressed as follows:

$$\rho_W S_v \Delta T = \rho_b Q_{es}/V_b + W_e - W_{ICE} + \rho_c \Delta \delta_{es} \qquad (19)$$

Relying solely on these variables cannot fully solve $Q_c(L)$ and $\Delta T$ because the number of unknown variables exceed the number of equations, it is necessary to explore the relationships in terms of volume. The gained or reduced cargo volume will be determined by the volume of the battery and the volume of the electric propulsion system compared to the volume of the ICE propulsion system:

$$V_c \Delta \delta_{es} = Q_{es} + V_e - V_{ICE} \qquad (20)$$

The right side of (20) represents the difference in space between the ICE ship and the ES. Assuming the average volume of a container is $V_c$, we can use equations (18)-(20) to solve $\Delta \delta_{es}$.

*B. Economic Model of the Trade-off*

Offshore charging will break the balance of this trade-off by changing the cruising range. The battery size will no longer be based on the total range but instead on the distances between charging stations along the route $L_v$, ensuring that the ship's battery size is sufficient to reach the next charging station as shown in (14). Therefore $L_v$ is a crucial variable to determine the objective function.

We analyze the results of different trade-off strategies from an economic perspective. First, considering the impact of cruising range on the economic aspects of battery size. The cost expression for battery size is given by (5), and cruising range affects the variables in (5), including battery size and the depreciation factor $\alpha_{dep,v}^{es}$. For battery size, according to (18)-(20), it is evident that a larger cruising range results in higher battery costs. However, the effect on the depreciation factor is more complex. The depreciation coefficient $\alpha_{dep,v}^{es}$ is calculated based on the battery's cycle count. Assuming the maximum cycle count of the battery is $N$, and the number of charging cycles during one voyage is $n$, and the number of charging cycles during one voyage can be calculated from the ship's cruising range, which is the ship's cruising range $L_v$ divided by the total distance traveled $L_{rt}$, rounded up. Then, the coefficient can be expressed as:

$$\alpha_{dep,v}^{es} = \frac{1}{N}\left\lceil \frac{L_{rt}}{L_v} \right\rceil \qquad (21)$$

The presence of the depreciation factor means that the battery cost does not decrease strictly in proportion to the reduction in battery size. However, this does not imply that there are no economic advantages, as the onshore electricity costs undergo proportional changes. The calculation of shore electricity costs is as follows:

$$C_{elec,v}^{es} = \mu_{reg}^{elec} Q_v^{es} \qquad (22)$$

where $\mu_{reg}^{elec}$ represents the unit price of shore electricity at different ports. Onshore electricity costs are directly proportional to battery size, so changes in cruising range will directly and proportionally affect shore electricity costs.

A decrease in cruising range reduces shore electricity costs but increases the cost of the OCS. The cost of the OCS is calculated as (2). The OCS also faces depreciation issues. Assuming the OCS has a lifespan of $W$ weeks, the total cost of the OCS is first distributed evenly across each week and then further allocated to each voyage within a week. The number of voyages corresponds to the average weekly shipping traffic volume. Therefore, the depreciation factor is expressed as follows:

$$\alpha_{depr,i}^{ocs} = \frac{1}{W \cdot \sum_{s \in S} w_s^{scn} \sum_{v \in D_s} v} \qquad (23)$$

The change in cruising range effectively shifts the onshore electricity costs to the OCS costs.

Changes in battery size will further drive the variation in cargo-carrying capacity. The quantity of cargo-carrying capacity is determined by constraint (14), and its revenue is calculated as follows:

$$C_{carg,v}^{es} = \mu_{carg} \Delta \delta_v^{es} \qquad (24)$$

where $\mu_{carg}$ represents the freight cost for a single TEU.

By using this trade-off approach, we can determine the most economically optimal combination.

V. CASE STUDY

In this section, we will analyze the coordinated planning of OCSs and ESs on the Shanghai-Busan route using AIS and GIS data.

*A. Data Description*

To account for varying geographical conditions across different regions and identify suitable installation sites, we accessed wind and solar radiation data from National Oceanic and Atmospheric Administration (NOAA)[19]. The NOAA collected wind and solar radiation data for the same route from 2020 to 2022 at 15×15-kilometer resolutions every 10 minutes. These datasets facilitate the evaluation of levelized cost of energy (LCOE) for potential offshore wind and solar power plants. Furthermore, General Bathymetric Chart of the Oceans (GEBCO)[20] provided high-precision ocean depth data at a resolution of 0.5×0.5 kilometers.

To gather authentic maritime data, we procured AIS data [21] for 714 container ships operating along the Shanghai-Busan route throughout 2022. This comprehensive dataset encompasses critical parameters such as departure and arrival times, voyage speeds, maximum velocities, cargo-carrying capacities, propulsion power, fuel tank volumes, and draft depths. These real-world maritime traffic patterns were utilized to formulate our OCS planning. The ships are classified into 8 clusters based on their cargo carrying-capacity in the unit of 20-foot equivalent unit (TEU). These 8 categories include: Small Feeder, Middle Feeder, Large Feeder, Wide-Beam, Classic Panamax, Small Neo-Panamax, Middle Neo-Panamax and Large Neo-Panamax.

*B. Result of Geographical Factor Analysis*

The optimization results of the geographical locations of the OCSs are shown in Fig.3. In Fig.3, 3 OCSs located at 221 kilometers, 456 kilometers, and 670 kilometers away from Yangshan Port in Shanghai. The distribution of OCSs are not evenly distributed as the geographical factors such as water depth and wind power are not identical along the route. However, the optimal position of the 3 OCSs are close to the evenly distributed point. It is because the main factor that determines the business case is the battery size of the ESs. The platform cost is relatively low compared to the total cost the OCS, as the water depth is shallow (as shown in Fig.5) and the wind power does not vary largely throughout the whole route, thus the position of each OCS is largely dependent on the maximum range of the ESs. It can be seen from Fig. 4, that if only 2 OCS are built, the battery cost of the ESs will be significantly higher to cover a longer range, which exceeds the cost reduction of building less OCSs, making total APC higher.

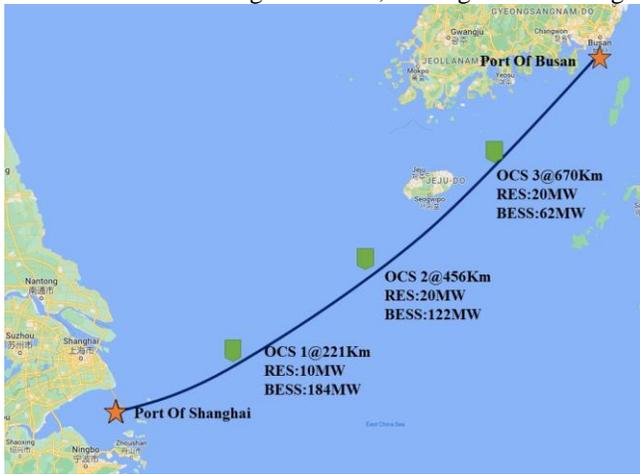

**Fig.3** The map displays the distribution and configuration of OCS on the Shanghai-Busan route.

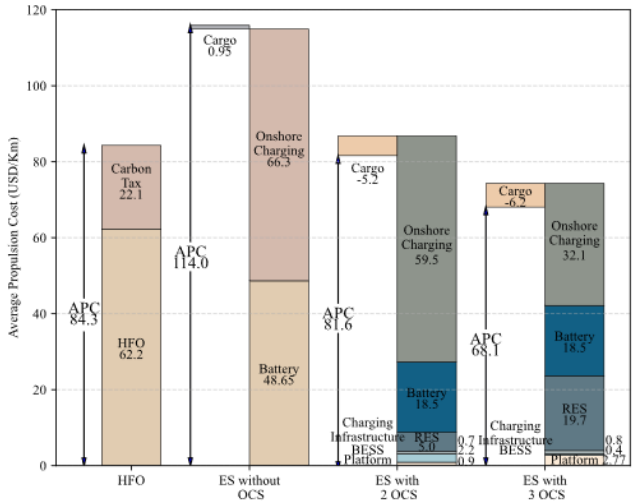

**Fig.4** Cost breakdown of HFO ships, ES, ES with 2 OCS and ES with 3 OCS.

The economic advantage of the OCSs can be also proved from the perspective of electricity price. To compare the electricity costs of RES and onshore electricity, the Levelized Cost of Energy (LCOE) of the OCSs can be calculated by dividing the average OCS cost by the average charged electricity in a single voyage. For average installation cost of OCS 1 to 3 is estimated 114 million US dollars and the total charging demand is approximately 4.76 billion MWh over 20 years. The LCOE of the OCSs of the Shanghai-Busan route is therefore $0.072/kWh, which is lower than the electricity prices for business of both China ($0.087/kWh) and South Korea($0.099/kWh)[22]. The reason is that the volume of between Shanghai and Busan is high, which effectively spread the installation cost over each charging activities.

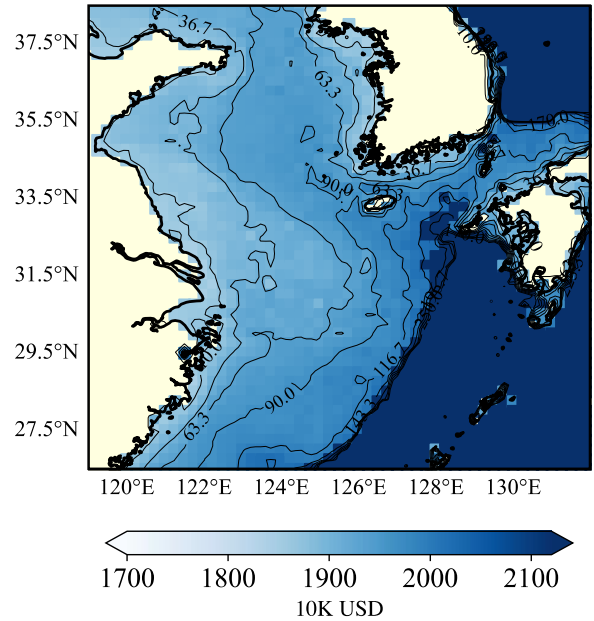

**Fig.5** The heat map of offshore platform costs in the China-South Korea region, along with the contour map of water depths.

*C. Economic Analysis of Trade-Off*

Based on the trade-off relationships of ESs (18)-(20), we conducted an analysis of the trade-offs for different ship types. Fig.6 displays the results of the Small Neo-Panamax container ship , taking into account the future changes in battery energy density. It is estimated that the energy density of batteries in the current year (2023), as well as in 2030 and 2050, are 300Wh/L, 700Wh/L, and 1200Wh/L, respectively [23].

The relationship between battery size and range forms a convex function, and an increase in battery energy density tends to make this relationship more linear. If the energy density of batteries increases in the future, this curve will be lower, meaning that the required battery size for the same cruising range will be lower. For example, on a range of 5000 km, the current battery density requires approximately 15834m$^3$ of battery size.

If the range is doubled, the battery size increases to about 36318 m$^3$, which is more than a doubling. Looking further ahead to 2050, it will be approximately 5733 m$^3$ and 12078 m$^3$, indicating a decrease of 63.8% and 66.7%, respectively. Therefore, increasing battery density can significantly reduce battery size while improving the linearity between cruising range and battery size.

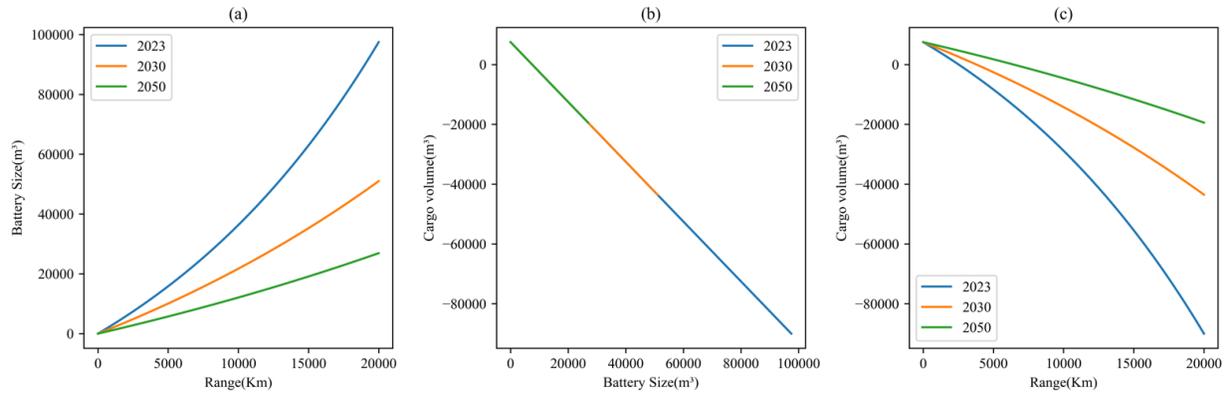

**Fig.6** The relationship between cruising range, battery size, and cargo-carrying capacity under different battery densities in current, 2023, 2030, and 2050 years.

As battery size increases, cargo-carrying capacity inevitably decreases. Fig.6(b) depicts the inverse relationship between cargo-carrying capacity and battery size, assuming that 76% of the change in cargo-carrying capacity can be used as the battery volume. Since the sum of cargo-carrying capacity and battery size remain constant, this relationship is linear.

The relationship between cruising range and cargo-carrying capacity is a concave function as shown in Fig. 6(c). It is evident that as the range increases, cargo-carrying capacity decreases. In the current scenario, for example, on a 5000-kilometer voyage from Qingdao to Singapore, a Neo-Panamax ship would experience a cargo-carrying capacity loss of 2.8%. If the journey extends further, such as from Guangzhou to Melbourne, covering approximately 10,000 kilometers, this proportion would increase to 9.8%. However, when the energy density of battery reaches 1200Wh/L by 2050, the Qingdao to Singapore route would actually gain a 0.6% cargo-carrying capacity, and the cargo-carrying capacity loss for the Guangzhou to Melbourne route would decrease to only 1.57%. Similar to Fig. 6(a), after the future increase in battery density, for the same cruising range, a greater cargo-carrying capacity can be achieved.

The changes in the ratio of the battery volume between ES and ES with OCS for 8 clusters of the ES on Shanghai-Busan route are shown in Fig. 6. It can be seen that all the ESs achieved a significant reduction in battery size with OCSs in the route. Specifically, Type 6 ES achieved the most significant reduction in battery size, reduced to about 1/4 of the original size. In contrast, Type 7 had the smallest reduction ratio, reduced to 32% of the original size The distributional impact on the battery size reduction is a direct consequence of the charging decision made by different ship types. Ships which have flexible arrival time windows will have more time to charge multiple times over the voyage, thus enjoy a more significant battery size reduction.

The use of OCS for ESs demonstrates exceptional economic advantages. Fig.4 also compares the economic performance of HFO ships, ESs without OCS, ESs with 2 OCSs and ESs with 3 OCS on the Shanghai-Busan route. ESs with 3 OCS exhibit the best economic performance, followed by HFO ships, while ESs without OCS have the lowest economic performance, with APC values of 84.3 USD/km, 91.3 USD /km, and 68.1 USD/km, respectively. It can be seen that currently ES itself cannot compete with HFO ships. OCS can significantly reduce the APC of the ES by 25%, even lower than the HFO ships by 19%.

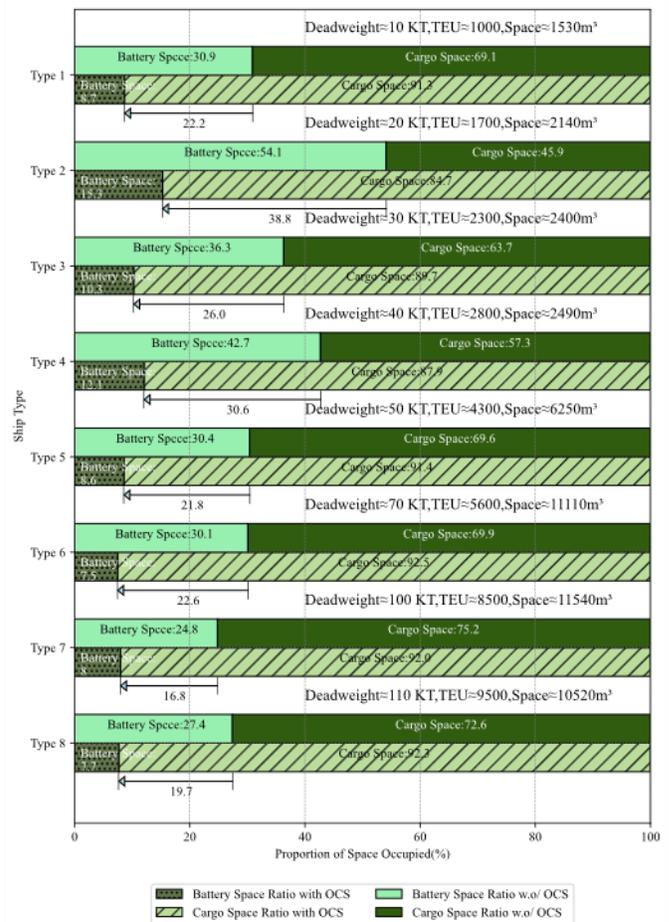

**Fig.7** The spatial distribution changes of eight ship types before and after using OCS.

This advantage is attributed to three main factors. Firstly, the reduction in battery size promotes a decrease in battery costs. As shown in Fig. 4, when not using OCS, the battery cost for ESs is 26.0 USD/km, whereas with OCS, this cost reduces to 18.5 USD/km. Secondly, the battery size is reduced from 2660 m$^3$ to 1681 m$^3$, 37% of the reduced space (979 m$^3$) can be used to carry 25 TEU more cargo. As shown in Fig.4, without OCS, the cargo revenue of ES is -0.95 USD/km compared with HFO ships, indicating that ES without OCS would require oversized batteries, resulting in a loss in cargo space and revenue. However, with the inclusion of offshore charging stations, cargo revenue is 6.2 USD/km compared with HFO, indicating that the reduced battery size frees up cargo space, leading to an increase in corresponding cargo revenue. Thirdly, OCS reduces the expensive onshore electricity costs. The cost of onshore electricity is higher due to the transmission cost, retail cost and carbon cost. Fig.4 shows a reduction of onshore electricity cost from 66.3 USD/km to 32.1 USD/km. Even considering the OCS charging cost, which is 23.7 USD/km, it still demonstrates a competitive edge.

Additionally, with stricter carbon tax policies in the future, the economic advantages brought by OCS may become more pronounced. Assuming that in 2030, carbon tax may increase to 163.9 USD/t, which implies that carbon tax costs will reach 40 USD/km. Even if battery technology doesn't advance further, or battery costs doesn't lower down, the APC of ES with OCSs is still likely to be superior to HFO fueled ships.

*C: Sensitivity Analysis*

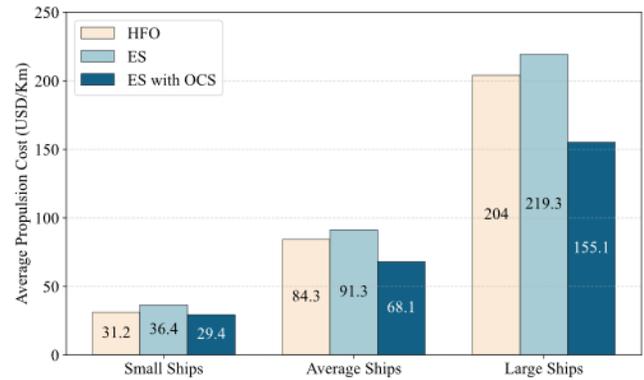

**Fig.8** The economic scenarios for HFO ships, ES, and ES with OCS are considered under 3 different shipping traffic volume: Small ship volume, Average ship volume and large ship volume.

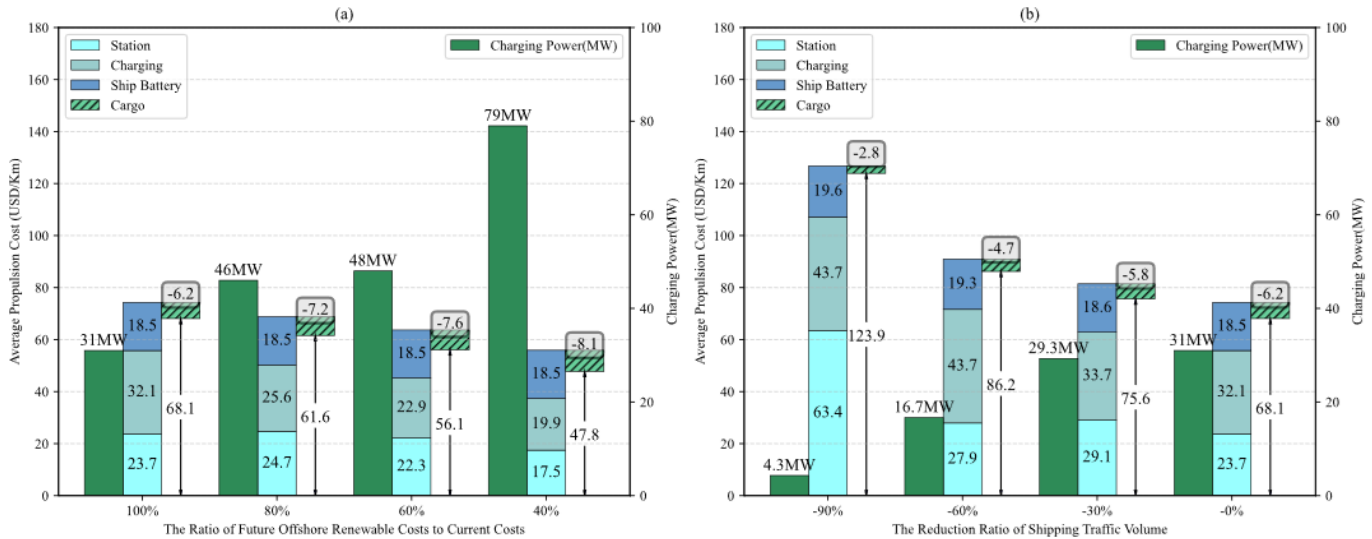

**Fig.9** Sensitivity analysis of OCS cost and shipping traffic volume. (a) When the cost of OCS decreases from the current (100%) to (40%), the change in the APC and charging power of OCS (b) The impact of a decrease in hipping traffic volume of -90% to no decrease on the APC and charging power of OCS.

*1)Ship Type Analysis:* Ships with higher propulsion power will benefit more from OCS. For the same range, they require larger battery installations, which can be effectively reduced by OCS. Generally, bulk cargo ships have lower propulsion power compared to container ships. To analyse the impact of fleet mix on our results, three scenarios are designed. The first scenario assumes a low-power bulk cargo ship dominated traffic. The second scenario mimic the current ship mix between Shanghai and Busan (18:82 between bulk and container). The third scenario a container dominated traffic. As shown in Fig.8, for scenario 1, the use of OCS (29.4 USD/km) is just slightly lower than HFO fueled ships, while for scenario 3, ES with OCS exhibit a 24% economic advantage, for the average scenario, a 19% economic advantage is gained by the ES. It can be seen that the traffic volume of the ESs between Shanghai and Busan has high energy demand, which is suitable for electrification.

*2)OCS Cost Analysis:* In the future, the cost of OCS is expected to continue decreasing. Key components of OCS costs include offshore platforms and wind turbines (or photovoltaic panels). According to statistics from Rystad Energy[23], the break-even prices for offshore platform projects are forecasted to continue declining, with deep water projects decreasing by 16% and shallow water projects by 10%. Regarding wind turbines, NREL predicts that by 2030, the LCOE of offshore wind turbine units will continue to decrease by 40% compared to 2016.

The decrease of the OCS cost will enhance the economic viability of OCS. Fig. 9(a) shows the APC of the ES when the OCS cost decreases to the 80%, 60% and 40% of the base price, it can be seen that decreasing the cost of OCS will not reduce the cost of OCS itself, but reducing the required battery size of ES by increasing the generation, energy storage and charging capacity of OCS. The reason is that the current charging capacity of OCS is not high enough to charge all of the ESs due to the high cost of OCS. If the cost of OCS decreases, enabling larger charging capacity of the OCS will lead to significant battery capacity of the ESs, as ESs get more chances to charge at OCSs. This benefit will outweigh the merely cost reduction of OCS itself.

*3) Shipping Traffic Volume Analysis:* Shipping traffic volume will alter the utilization of OCS, thereby altering the APC. For the same OCS, the more ES participate in charging, the higher the utilization and lower cost spreading of OCS. However, ES will not reach a very high occupancy rate in a short time, as the existing HFO fueled ships is still on active service. It is valuable to estimate the APC of ES with the gradually increase of electrification rate. Fig. 9(b) shows the APC of the ESs and generation/charging capability of the OCSs in different ES traffic volume between Shanghai and Busan. With the increase of the ES traffic volume, the overall APC constantly decreases. When looking into the cost breakdown, it can be seen that the station cost experience a hump. The reason is that when the total ES traffic volume is very low, the spreading of the OCSs will gradually decrease with the volume increases, when the traffic volume increases to a certain extent, more RES, BESS and charging devices are required to provide larger charging capacity, thus the required battery size of the ESs could be reduced (as mentioned above), which leads to better total APC. When the ES traffic volume reaches 40%, the ES will achieve the cost parity with HFO fueled ships. The above result can be served as a reference to the maritime policy makers, that similar subsidy policies like electric vehicles can be made for ES when the electrification rate of ES is not high at the early stage.

VI. CONCLUSION

This paper introduces a method for the coordinated planning of OCSs and electrified ships, with the goal of minimizing average propulsion costs. It takes into account the variations in location due to the complex maritime geographical environment, marking the first practical application of site selection and capacity planning for OCSs. Through this method, specific planning proposals for OCSs tailored to particular routes can be developed, reducing the economic losses associated with electrification.

By studying the impact of geographical factors and shipping traffic volume on planning, as well as the trade-offs in configuring ESs, two challenging aspects of the planning model have been addressed. Leveraging AIS and GIS data, our method has been applied to the Shanghai to Busan route. The results indicate that OCSs have a significant effect on reducing average propulsion costs. Additionally, type of ships, offshore renewable costs, and shipping traffic volume significantly influence the planning.

While the conclusions are currently limited to the Shanghai-Busan route, the economic benefits of OCSs can evidently be extended to other routes, furthering the electrification of routes with similar conditions.